\documentclass{vgtc}                          

\graphicspath{{figures/}{pictures/}{images/}{./}} 

\usepackage{times}                     

\usepackage{tabu}                      
\usepackage{booktabs}                  
\usepackage{lipsum}                    
\usepackage{mwe}                       

\usepackage{mathptmx}                  

\onlineid{0}

\vgtccategory{Research}

\vgtcinsertpkg

\title{Visualization Was Here: Reorienting Research When Visualizations Fade into the Background}

\author{Paul C. Parsons\thanks{e-mail: parsonsp@purdue.edu}\\ %
        \scriptsize Purdue University} %

\abstract{
Visualization research often centers on how visual representations generate insight, guide interpretation, or support decision-making. But in many real-world domains, visualizations do not stand out—they recede into the background, stabilized and trusted as part of the everyday infrastructure of work. This paper explores what it means to take such quiet roles seriously. Drawing on theoretical traditions from joint cognitive systems, naturalistic decision making, and infrastructure studies, I examine how visualization can become embedded in the rhythms of expert practice—less a site of intervention than a scaffold for attention, coordination, and judgment. I illustrate this reorientation with examples from mission control operations at NASA, where visualizations are deeply integrated but rarely interrogated. Rather than treat invisibility as a failure of design or innovation, I argue that visualization’s infrastructural presence demands new concepts, methods, and critical sensibilities. The goal is not to diminish visualization’s importance, but to broaden the field’s theoretical repertoire—to recognize and support visualization-in-use even when it fades from view.
} 


\begin{document}



\maketitle

\section{Introduction}

Visualization research often centers on how visual representations can amplify human insight, support interpretation, and enable better decisions. But in many settings, visualizations do not stand out. They do not dazzle, surprise, or provoke. Instead, they settle quietly into the background—used continuously, trusted implicitly, and rarely discussed. In such contexts, visualization is no longer a tool for exploration or persuasion. It becomes infrastructure.

This paper explores what it means to take visualization’s infrastructural roles seriously. My argument is grounded in a moment of dissonance that emerged during a study involving NASA’s mission control operations—an environment saturated with visualizations, yet structured around verbal coordination, procedural routines, and tacit judgment. This experience prompted a broader rethinking of visualization’s role in expert practice. These visualizations are crucial to operations, yet they are rarely the focal point of decision-making. They are referenced, glanced at, or ambiently monitored—subordinated to verbal coordination, procedures, scripts, and interpretive judgment. In this setting, visualization plays a supporting role in joint cognitive work: stabilizing attention, cuing action, and scaffolding collective understanding without demanding it.

Such contexts challenge how visualization researchers typically engage with real-world practice. They prompt a shift in orientation: from treating visualization as an intervention to recognizing it as embedded support; from valuing design novelty to attending to infrastructural presence; from seeking moments of insight to tracing rhythms of coordination. This shift draws on traditions such as joint cognitive systems (JCS) \cite{woods_joint_2006, hollnagel_joint_2005}, naturalistic decision making (NDM) \cite{klein_decision_1993}, and infrastructure studies—including domestication theory~\cite{mansell_design_1996} and Star’s work on invisible work and maintenance~\cite{star_steps_1994}. It also aligns with recent perspectives on embedded and situated visualizations, which emphasize how representations become woven into physical environments, cognitive routines, and social interactions~\cite{willett_embedded_2017, bressa_whats_2022, elmqvist_ubiquitous_2013}.

This paper calls for a reorientation: What becomes of visualization when it is no longer novel? What becomes of visualization research when its object of study has already been absorbed into practice? These questions are motivated by a moment of conceptual dissonance—when the assumptions I brought as a visualization researcher came into tension with the realities of embedded use. 

The provocation offered here is directed toward visualization researchers studying professional, high-context environments—places where visualizations are used not by casual viewers or the general public, but by trained practitioners embedded in complex systems of coordination. In such contexts, the assumption that visualization is central, insight-generating, or intervention-worthy may no longer hold. This is not to say visualization is unimportant. Rather, its importance may lie elsewhere---e.g., in how it quietly stabilizes attention, cues action, or supports joint judgment. By calling attention to these roles, this paper aims to reorient how we study, theorize, and participate in visualization practice in domains where visualization has already been absorbed into the background of expert work.


This paper invites a rethinking of visualization’s role in mature, high-context domains. By reframing visualization as a quietly embedded and infrastructural component of expert work, it shifts the emphasis away from novelty, salience, and intervention. It surfaces methodological and conceptual challenges that arise when visualizations fade into the background of practice, raising questions about how to recognize, study, and support visualization-in-use when it is already stabilized and domesticated. Drawing on perspectives from infrastructure studies, cognitive systems engineering, and naturalistic decision making, the paper calls for an expansion of the field’s theoretical repertoire and encourages greater attention to judgment, coordination, and the sociomaterial conditions that shape visualization’s enduring presence in complex work systems.

\section{Visualization, Already Embedded: A Reorientation in Practice}

This paper began with a moment of conceptual dissonance. As a visualization researcher working on a project exploring smart space habitats, I expected to study and (re)design data visualizations used to improve monitoring, anomaly detection, and decision-making. The environment seemed well suited for such work: vast telemetry streams, constant monitoring, high-stakes decisions, and extensive use of visual interfaces across multiple screens. But what I encountered challenged these assumptions. Visualizations were everywhere, but they were typically not the focus. They were stable, simple, and largely taken for granted. Flight controllers did not scrutinize them for patterns or insights. Instead, they monitored, scanned, and coordinated using a mix of displays, verbal communication, procedural scripts, and embodied expectations. Instead of visualization acting as an intervention to be introduced, it was an element of practice that had already settled into infrastructure.

This realization prompted a reorientation in my approach. The familiar frameworks—design studies, evaluation protocols, and distributed cognition—offered little guidance for analyzing visualization in this setting. What was needed was not a better visualization, but a better way of understanding visualization’s embedded role: how it participates in the rhythms of expert work, supports anticipation, and scaffolds coordination—without ever drawing attention to itself. To make sense of this role, I turn to perspectives that foreground the quiet, backgrounded functions of artifacts. Cognitive systems engineering (CSE) and naturalistic decision making (NDM) offer models of joint cognitive systems, where people and tools adapt together under constraint. Infrastructure studies and domestication theory emphasize how technologies become invisible—not because they fail, but because they work. These frameworks offer ways to think about visualization not as a spectacle or discovery tool, but as a substrate for skilled action and shared judgment.

\section{Revisiting Collaboration in Visualization}

Collaboration and multidisciplinary work are deeply ingrained in the foundations of visualization research. The field has long understood its impact to be amplified through partnerships with domain experts, and has built methodological frameworks that emphasize the importance of iterative, human-centered collaboration. Early commentaries (e.g., McCormick~\cite{mccormick_visualization_1987}) and concepts like Donna Cox’s ``Renaissance Team''~\cite{kirby_visualization_2013} articulated a vision of visualization as a deeply interdisciplinary activity, with contributions shaped by the needs of partner domains. Contemporary design frameworks, like the Design Study Methodology~\cite{sedlmair_design_2012}, have institutionalized these collaborative ideals, offering structured processes for engaging in collaborative visualization development.

These frameworks continue to provide valuable structure, though they are often oriented around particular assumptions about the purpose and trajectory of collaboration. Most notably, they center the visualization artifact as the locus of innovation and intervention. Collaboration is typically framed as a process through which visualization researchers identify user needs, co-design representations, and deploy solutions. In this model, visualization serves as a lever—something introduced into the domain to address a gap or enhance understanding.

However, this artifact-centric view becomes strained in contexts where visualizations are already integrated into practice. As our field sites and applications mature, researchers will increasingly encounter domains where visualizations are not novel contributions, but longstanding elements of the working environment. In such cases, the collaborative task is not to design new artifacts, but to understand how existing ones participate in ongoing coordination, decision-making, and sensemaking. The epistemic role of visualization is less about generating insight and more about sustaining shared expectations, stabilizing attention, and scaffolding routine work.

Recent scholarship has begun to surface similar tensions. Akbaba et al.~\cite{akbaba_troubling_2023} critique the ethical and epistemic asymmetries embedded in visualization design studies, highlighting the under-acknowledged labor of care, maintenance, and relationship-building that sustains collaborative work. Meyer and Dykes~\cite{meyer_criteria_2020} raise concerns about the “ethics of exit,” questioning what researchers leave behind after deploying visualization systems and how responsibility for long-term upkeep is distributed. Xing et al.~\cite{xing_collaborating_2025} emphasize the fluidity of roles in collaborative visualization work, arguing that project participants often traverse boundaries between design, domain knowledge, and technical implementation, sometimes holding multiple responsibilities at once. These studies suggest collaboration is not a fixed arrangement of complementary expertise, but a dynamic process of role negotiation, institutional navigation, and infrastructural entanglement.

Several accounts add nuance the traditional division of roles in collaborative studies. For instance, Kirby and Meyer~\cite{kirby_visualization_2013} identified patterns of successful collaboration, such as deep mutual engagement and shared learning, while also cautioning against partnerships in which visualization is treated as a service rather than as co-inquiry. Hlawitschka et al.~\cite{chen_collaborating_2020} called for more structured and continuous cross-disciplinary dialogue, emphasizing how early and sustained interaction fosters mutual understanding. Similarly, Keefe’s work on immersive tools for scientists and artists~\cite{keefe_understanding_2014} challenged assumptions about utility and intention, showing how visualizations designed for one purpose often catalyze unexpected forms of collaboration and discovery.

These insights reveal collaboration as an emergent, situated phenomenon—one that unfolds not just through co-design, but through shared engagement with visual artifacts over time. In domains such as mission control, clinical diagnostics, or emergency response, collaboration is not about producing new visualizations. It is about managing representational ecologies in which visualizations are already present, already trusted, and already part of the lifeworld of professional practice. In these contexts, to collaborate is to coordinate around infrastructure—not to design it anew. These tensions echo a sentiment voiced two decades ago by Lorensen~\cite{lorensen_death_2004}, who argued that visualization, having achieved many of its early ambitions, risked fading into irrelevance through quiet assimilation into other disciplines and infrastructural functions. His provocation serves as a reminder that perspectives on collaboration can evolve alongside other aspects of the field. In domains where visualization is already embedded, the most impactful contribution may not be new artifacts, but new understandings of how those artifacts function within sociotechnical systems.

My perspective shares similarities with prior efforts to reconceptualize the collaborative roles of visualization in embedded settings. Willett et al.~\cite{willett_embedded_2017} propose embedded data representations as a way to understand how visualizations support task completion in place, often through peripheral awareness rather than direct manipulation. Bressa et al.~\cite{bressa_whats_2022} survey perspectives on situated visualization and emphasize the importance of context, embodiment, and interpretive practice. Elmqvist and Irani~\cite{elmqvist_ubiquitous_2013} similarly argue for “ubilytics” as an analytic lens for ubiquitous analytics, where representations are dispersed and woven into the everyday. These works suggest that collaboration is not only about designing tools together—it is about inhabiting shared representational landscapes and making sense of them in context.


\section{Studying Visualization-in-Use in Mission Control}

This perspective is grounded in a qualitative study of NASA flight controllers involved in Shuttle and International Space Station operations. While full methodological details are reported elsewhere (see \cite{zhang_designing_2023, murali_krishnan_habsim-hms_2024, zhang_visual_2022, parsons_adaptive_2022}), this material provides the basis for the interpretive insights presented here. These experts are responsible for monitoring and managing the International Space Station (ISS), which produces dozens of terabytes of telemetry data each day. Within the Mission Control Center, this data is visualized across dozens of screens. Each controller typically works across four to six monitors, with displays tuned to different subsystems and timescales. These representations are essential for situational awareness, yet they are not complex or flashy. Their visual grammar is deliberately simple. The predominant display formats include tables, two-variable line graphs, color-coded status indicators, and text-based logs. These visualizations do not rely on novel encodings or advanced interactive features. Instead, they are designed for clarity, consistency, and speed of interpretation. Their simplicity is not a constraint but a form of adaptation—reflecting the need for stability, redundancy, and interpretive fluency in a high-consequence environment.

In interviews, flight controllers described two primary modes of engagement with these displays: passive monitoring and active scanning. Passive monitoring involves maintaining broad situational awareness, with visual attention triggered by salient cues such as a color change or threshold breach. Active scanning, by contrast, entails rhythmic, routinized sweeps across displays to detect subtle deviations from expected patterns. These strategies support a deeply anticipatory form of reasoning—one rooted in experience, expectation, and embodied familiarity. This mode of interpretation depends not only on the displays themselves, but also on how they are configured and personalized. Controllers often adapted their visual environments over time—developing idiosyncratic scan paths, modifying layout preferences, and even annotating screens with masking tape or post-it notes to delineate risk zones or command boundaries. These acts of customization were not formally taught. They were acquired through mentorship, observation, and repeated engagement with the tools. In this sense, visualization was not a static interface, but a substrate for situated design—malleable, lived-with, and shaped in practice.

Crucially, visualizations did not serve as the sole source of evidence or decision input. Interpretation was distributed across multiple forms of representation, roles, and communication channels. Controllers coordinated verbally, consulted procedural checklists, referenced historical logs, and leveraged shared mental models to make sense of evolving situations. Visualizations anchored this coordination, but rarely dictated it. Their role was to cue attention, stabilize interpretation, and provide a common substrate for joint sensemaking.

These patterns resonate with the notion of visualization as part of a joint cognitive system~\cite{hollnagel_joint_2005, woods_joint_2006}—a framework rooted in cognitive systems engineering that emphasizes how people and artifacts work together as a system to maintain performance under pressure. Unlike general accounts of distributed cognition, which describe how cognitive tasks are spread across tools and agents, the JCS perspective foregrounds coordination, adaptation, and resilience in high-stakes environments. In this view, visualizations do not stand apart as instruments of analysis. Instead, they recede into the fabric of professional practice, contributing not through salience or novelty, but by quietly supporting fluent action, shared mental models, and institutionalized forms of judgment.

For visualization researchers, this raises a methodological and conceptual challenge. Much of the field prioritizes visual engagement, interpretability, and insight generation. But in domains like mission control, visualization succeeds precisely when it disappears—when it supports expert performance without demanding attention. Studying such systems requires a shift in emphasis: from evaluating visual artifacts in isolation to understanding how they function within sociotechnical systems, maintain fluency under pressure, and become infrastructural to expert work.

\section{Theoretical Foundations for Studying Embedded Systems}

Having briefly described how visualization functions in the embedded, high-stakes environment of mission control, I now turn to broader theoretical perspectives that help make sense of these dynamics. While the focus in this paper is ultimately on visualization, the patterns observed—routine use, tacit interpretation, ambient monitoring—are not unique to visual representations. They are features of how many technologies become integrated into the everyday infrastructure of expert work.

Several theoretical traditions offer conceptual tools for understanding such embeddedness. These include infrastructure studies, domestication theory, joint cognitive systems (JCS), and naturalistic decision making (NDM). Though distinct in focus, they share a common orientation: that technology’s significance often lies not in visibility or novelty, but in how it supports situated action, shared attention, and resilient coordination. These perspectives help reframe visualization not as an interventionist tool, but as a participant in distributed systems of activity and meaning.

\textbf{Infrastructure Studies.} Star and Bowker famously argue that infrastructure is most visible when it fails~\cite{star_steps_1994, bowker_sorting_2000}. Infrastructures are embedded, relational, and often invisible—revealed not through their presence, but through breakdowns, repairs, and workarounds. These insights have shaped scholarship in science and technology studies (STS) and human-computer interaction (HCI), where researchers increasingly study the mundane, the residual, and the maintained. Applied to visualization, this perspective shifts attention away from novel representations and toward how visualizations are stabilized, depended upon, and maintained as part of broader sociotechnical systems.

\textbf{Domestication Theory.} Originating in media and technology studies, domestication theory describes how new technologies are integrated into everyday life through processes of appropriation, normalization, and routinization~\cite{mansell_design_1996}. Once novel, technologies become backgrounded—taken for granted and folded into routine practice. Domestication offers a lens for thinking about visualization not as a disruptive intervention, but as a settled component of organizational routines. It draws attention to how visualizations are made familiar, how they fade from awareness, and how they become invisible through stable use.

\textbf{Joint Cognitive Systems (JCS).} JCS and CSE focus on how people and artifacts form integrated systems of cognition and action~\cite{hollnagel_joint_2005, woods_joint_2006}. Representations—visual, textual, procedural—are not external aids, but active components of the cognitive system. This perspective emphasizes coordination, adaptation, and the situated use of representations to support joint activity. In visualization contexts, it suggests that visualizations may not stand alone as sites of reasoning, but instead function as part of distributed representational ecologies that scaffold interpretation, cue responses, and maintain common ground.

\textbf{Naturalistic Decision Making (NDM).} NDM investigates how experts make decisions in real-world settings characterized by uncertainty, time pressure, and complexity~\cite{klein_decision_1993}. Instead of following normative models, expert decision-makers rely on experience, recognition-primed strategies, and sensemaking grounded in context. Visualizations in these settings often serve as peripheral cues or tools for maintaining situational awareness—not as engines of insight, but as background supports. This perspective highlights the non-linear, contingent, and judgment-laden nature of visualization use in practice.


\section{Rethinking Visualization’s Role through Alternative Theoretical Lenses}

The perspectives outlined in the previous section foreground \textit{invisibility}, \textit{routine}, and \textit{joint activity} as defining characteristics of mature work systems. But what do these insights mean for visualization research? This section turns inward to examine how such interdisciplinary frameworks challenge common assumptions in the field. In particular, they invite us to reconsider the idea that a visualization’s value lies in \textit{novelty}, \textit{sophistication}, or \textit{the capacity to provoke insight}. Instead, they suggest alternate criteria: \textit{sustained relevance}, \textit{quiet integration}, and \textit{support for coordination under pressure}.

These frameworks offer conceptual tools for reorienting how we study, design, and evaluate visualization in professional, high-context environments. From \textit{Cognitive Systems Engineering (CSE)} and \textit{Naturalistic Decision Making (NDM)}, I adopt the view that expert work is rarely about optimizing discrete decisions. Instead, it involves navigating uncertainty, anticipating anomalies, and managing competing priorities in fluid, constrained conditions. Visualizations in such contexts function as components of a \textit{representational ecology}—not as standalone interfaces for discovery or persuasion, but as scaffolds for \textit{judgment} and \textit{coordination}. The concept of the \textit{joint cognitive system}~\cite{woods_joint_2006} captures this orientation: people and artifacts operate as adaptive systems, working together to sustain resilience. Our recent work~\cite{parsons_judgment_2025} extends this view into design practice, framing visualization design itself as a form of \textit{coordinated judgment} within distributed sociotechnical systems.

In parallel, theories from \textit{domestication}~\cite{mansell_design_1996} and \textit{infrastructure studies}, especially Star’s work on invisible work~\cite{star_steps_1994}, highlight how representations, once absorbed into use, may \textit{recede from awareness}—not because they are unimportant, but because they have become essential. Visualization may succeed not by attracting attention, but by \textit{quietly stabilizing the conditions of work}.

To some extent, the visualization literature has already gestured in this direction. Liu et al.~\cite{Liu2008} proposed \textit{distributed cognition} as a framework for understanding how visual artifacts support coordinated activity across tools and actors. Willett et al.~\cite{willett_embedded_2017} introduced a taxonomy of \textit{embedded data representations}, showing how visualizations can become spatially, temporally, or socially entangled in activity. Bressa et al.~\cite{bressa_whats_2022} elaborated on this with a framework for \textit{situated visualization}, emphasizing how data representations embed into action, space, and routine. Similarly, Elmqvist and Irani~\cite{elmqvist_ubiquitous_2013} framed \textit{ubiquitous analytics} as lightweight, ambient, and peripherally accessible visualizations that fade into background activity.

These contributions offer a valuable descriptive vocabulary for characterizing embedded visualization. Yet most stop short of explaining how these representations function as part of complex, evolving systems of expert coordination. They help us observe \textit{what} embedded visualization looks like—but offer fewer tools for understanding \textit{how} such representations support resilience, adaptation, or system-level sensemaking. For that, I turn to CSE, NDM, and infrastructure studies, which bring with them robust theoretical foundations for analyzing work under uncertainty, breakdown, and constraint. These frameworks are not offered as replacements, but as necessary supplements—especially in domains where visualization’s power lies in its invisibility.

Together, these perspectives suggest that visualization’s impact in complex settings may lie not in \textit{intervention}, but in \textit{integration}. This shift—from \textit{designed insight} to \textit{embedded support}—requires rethinking how we study, theorize, and participate in visualization practice.

\subsection{Fluency: When Visualization Use Becomes Invisible}

One concept from CSE that resonates with these observations is the \textit{Law of Fluency}~\cite{woods_joint_2006}. It describes how expert performance in well-adapted joint cognitive systems often appears smooth and effortless—masking the complexity, improvisation, and skilled judgment required to sustain it. In mission control, the use of visualizations exemplifies this law: displays are scanned and referenced fluidly, yet this fluency hides the tacit knowledge, scan strategies, and adaptive reasoning that flight controllers cultivate over time. The smoother the work appears, the harder it is to notice the cognitive scaffolding that makes it possible.

The Law of Fluency underscores a core challenge: once visualizations become fully integrated into practice, they become harder to study. Their role is no longer to prompt insight, but to quietly sustain performance under pressure. Researchers seeking to understand visualization-in-use must shift their attention—from visible interactions to hidden adaptations, from designed features to emergent functions, from novelty to normality. More broadly, the Law of Fluency belongs to a family of \textit{Laws of Adaptation} in CSE that describe how work systems evolve to meet the demands of surprise, uncertainty, and constraint. It reminds us that expertise is not only what practitioners know, but how they adapt that knowledge in context. And it suggests that visualization, when fully domesticated, becomes not a lens for seeing data—but a means of \textit{seeing smoothly}.

\subsection{Plural Perspectives on Embedded Visualization}

While this paper foregrounds concepts from CSE and NDM, these are not the only lenses available. My emphasis on decision-making, coordination, and adaptation shaped the theoretical orientation of this work—but other framings could illuminate different facets of backgrounded visualization. For instance, infrastructure studies and STS emphasize the political, affective, and material dimensions of mundane tools. A visualization may become infrastructural not because it supports fluency, but because it affords comfort, familiarity, or aesthetic pleasure. In such contexts, invisibility might reflect \textit{habituation}, \textit{attachment}, or even \textit{ritualized play}, particularly in domains like education, personal informatics, digital journalism, or art. Cultural studies, design theory, and media studies offer further vocabulary for understanding these dynamics—terms like \textit{ambience}, \textit{aesthetic friction}, \textit{domestication}, \textit{situated action}, and \textit{practice ecologies}. These approaches complement cognitive framings by attending to emotional and experiential qualities that may also underpin infrastructural invisibility.

By drawing primarily from cognitive and systems-oriented theories, I offer one perspective on visualizations that fade into the background. Others are possible—and necessary. Invisibility is not always about fluency or resilience. Sometimes, it is about trust, comfort, or the passage of time.

\section{Summary}

This paper has offered a provocation: in many expert domains, visualization no longer stands out as novel. It becomes infrastructure—quiet, trusted, and woven into the rhythms of professional practice. In contexts like NASA mission control, visualizations support judgment and coordination not by prompting insight, but by stabilizing attention and sustaining shared expectations. Such settings complicate conventional research approaches. Methods focused on salience or usability may miss how visualizations operate once they are embedded. To study these roles, researchers need more than new theories; they need ways of noticing what is usually invisible—how visualizations cue action, support personalization, and anchor cross-representational coordination. Perspectives from infrastructure studies, naturalistic decision making, and cognitive systems engineering can help, not as prescriptions but as starting points for this shift. The challenge is to develop practical habits of inquiry that surface the subtleties of embedded visualization—its stability, fluency, and maintenance—without re-centering novelty. To remain relevant in high-context domains, visualization research must be willing to follow visualization where it goes, even when that means shifting our gaze from what is striking to what is quietly at work in the background.


\bibliographystyle{abbrv-doi}

\bibliography{references}
\end{document}